# Propagation of Spin Waves in Doubly Periodic Magnonic Crystals

Adam G. Whitney[1], Joshua M. Lewis[2], Justin Dickovick[3], Vijay Kalappattil[3], Lincoln D. Carr[2], and Mingzhong Wu[3,4,5,1*]

[1]*Department of Physics, Colorado State University, Fort Collins, Colorado 80523, USA*
[2]*Department of Physics, Colorado School of Mines, Golden, Colorado 80401, USA*
[3]*Department of Physics, Northeastern University, Boston, Massachusetts 02115, USA*
[4]*Department of Electrical and Computer Engineering, Northeastern University, Boston, Massachusetts 02115, USA*
[5]*Quantum Materials and Sensing Institute, Northeastern University, Burlington, Massachusetts 01803, USA*

Towards the development of strategies for tailoring spin-wave band gaps in magnonic crystals, this work examines the band gap properties in a one-dimensional magnonic crystal with double periodicity. A long and narrow yttrium iron garnet (YIG) thin film strip is etched with an array of transverse groove lines separated by alternating distances, where the second distance is twice the first. This double periodicity in the magnonic crystal translates into dissimilar band gaps in the frequency domain, with the third and sixth band gaps being more pronounced than others. These band gaps are more pronounced because the corresponding wavenumbers simultaneously satisfy the Bragg scattering conditions for the periods equal to the two groove separations as well as their sum. Experimental observations are reproduced by numerical simulations. Together, the experimental and numerical results demonstrate how multiple periodicities could be an effective design parameter for creating magnonic crystals with desired band gaps.

## I. Introduction

A magnonic crystal is an engineered magnetic material with periodically modulated magnetic properties. These spatial periodic variations give rise to Bragg reflections or scattering and thereby create band gaps for spin waves in the material. These band gaps occur at frequencies corresponding to the Bragg wavenumbers — the wavenumbers at which Bragg reflection occurs. The rich features associated with these band gaps have facilitated the study of intriguing physics that are otherwise inaccessible with spin waves in continuous media. Examples include the experimental observation of spin-wave soliton fractals in one-dimensional (1D) magnonic crystals[1] and the theoretical prediction of topological magnons in two-dimensional (2D) magnonic crystals.[2,3]

Previous work on 1D magnonic crystals has predominantly focused on structures with single periodicity.[1,4,5,6,7,8,9,10] The bandgaps in such materials exhibit similar properties, as expected according to Bragg's law. In an effort to develop strategies for tailoring band gaps, this study examines the band gap properties in a 1D magnonic crystal with double periodicity. A long and narrow yttrium iron garnet (YIG, $Y_3Fe_5O_{12}$) thin film strip is etched with an array of transverse line grooves whose separation alternates between $a_1$ and $a_2 = 2a_1$. This double periodicity in the spatial domain translates into dissimilar transmission dips in the frequency domain, with the third and sixth transmission dips being broader and deeper than others. These results indicate that every third band gap is more pronounced. The underlying reason is that the wavenumbers corresponding to those band gaps simultaneously satisfy the Bragg conditions for the periods $a_1$, $a_2$, and $a_1 + a_2$. The experimental observations are reproduced by simulations using a transfer matrix method. The experimental and numerical results together demonstrate an effective strategy for creating magnonic crystals with desired band properties for both fundamental research and device applications. From a fundamental perspective, for example, magnonic crystals with broad band gaps ease the experimental investigation of in-gap modes. Practically, magnonic crystals with varying band gaps can enable easy switching between different types of microwave filtering operations.

*min.wu@northeastern.edu

## II. Sample Preparation

This study made use of three samples: (1) an un-patterned or uniform YIG thin film strip, (2) a singly periodic magnonic crystal, and (3) a doubly periodic magnonic crystal. The magnonic crystals were created through etching arrays of transverse line grooves on the top surfaces of long and narrow YIG thin film strips. For all the samples, the YIG films were grown on (111)-oriented single-crystal gadolinium gallium garnet (GGG, $Gd_3Ga_5O_{12}$) substrates by liquid phase epitaxy. The fabrication of the line grooves involved pattern formation through a laser writing-based lithography process (using AZ 1512 photoresist) and the chemical etching of exposed YIG surfaces by hot acid.

The exposed sections of the YIG strips were chemically etched by submerging the samples in hot $H_3PO_4$. The etching rate is controlled by varying the acid temperature and was approximately 10 nm/s at 160 °C, which is comparable with values reported previously.[6,9] The temperature uniformity of the acid is critical for an even etching. To ensure good temperature uniformity, the beaker was submerged in a bead bath of alumina pellets. After etching, samples were cut to desired dimensions using a diamond wire saw and the photoresist was then removed in an acetone bath using an ultrasonic cleaner.

Table 1. Properties of three samples: a uniform YIG thin film strip, a singly periodic YIG-based magnonic crystal, and a doubly periodic YIG-based magnonic crystal.

| Properties | Uniform YIG strip | Singly periodic magnonic crystal | Doubly periodic magnonic crystal |
|---|---|---|---|
| Sample width (mm) | 1.91 | 1.99 | 1.98 |
| Sample length (mm) | 17.4 | 17.8 | 22.6 |
| Film thickness (μm) | 10.5 | 12.7 | 4.2 |
| Groove count | - | 31 | 40 (20 pairs) |
| Groove spacing (μm) | - | $400 \pm 4$ | $a_1 = 149 \pm 10$<br>$a_2 = 300 \pm 12$ |
| Groove depth (μm) | - | $1.16 \pm 0.02$ | $0.84 \pm 0.03$ |
| Groove width - top (μm) | - | $103 \pm 2$ | $65 \pm 15$ |
| Groove width - bottom (μm) | - | $42 \pm 2$ | $27 \pm 1$ |
| Saturation induction (G) | 1898 | 1926 | 1790 |

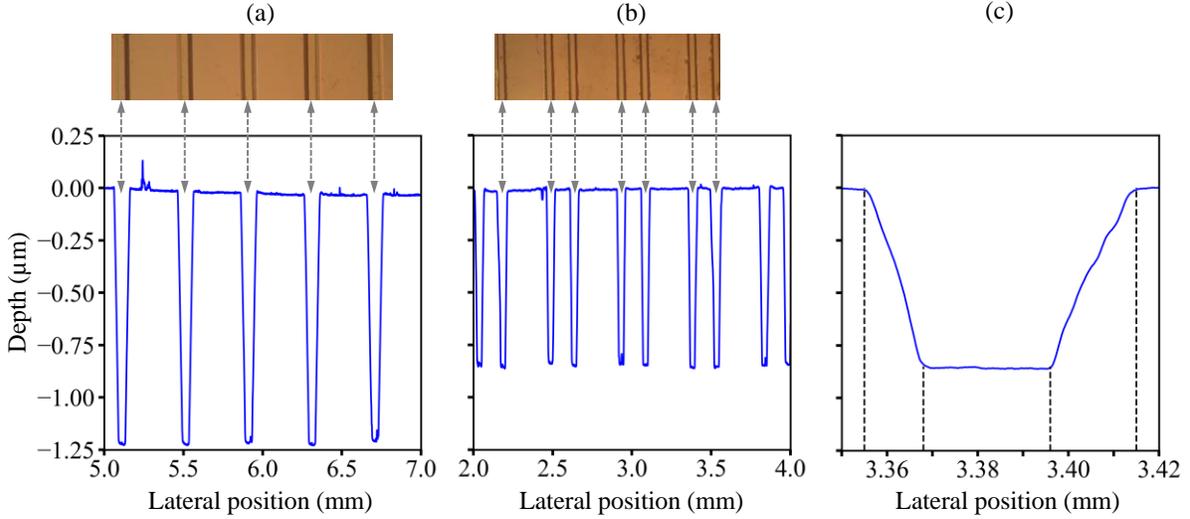

Fig. 1. Profilometer measurements of the two etched YIG strips across a length of 2.0 mm. (a) Depth profile of a singly periodic magnonic crystal that shows five grooves with depths of about 1.16 µm. The slight slope of the top surface is due to a small misalignment of the profilometer stage. (b) Depth profile of a doubly periodic magnonic crystal that shows five pairs of grooves with depths about 0.84 µm. The images on the top are the corresponding microscope pictures of the samples. (c) Expanded view of the 7$^{th}$ groove in (b).

Table 1 lists the properties of the three samples. The groove properties were determined through surface profile measurements using a Bruker Dektak XT stylus profilometer. Figure 1 shows a portion of the surface profile together with an optical microscope image for each magnonic crystal sample. The singly periodic sample has a lattice constant $a = 400$ µm that is defined as the distance between the falling edge of one groove to the falling edge of the next. There are 31 identical grooves in total, spanning a total length of 12.1 mm. Each groove is wider at the top than at the bottom, as can be seen in Fig. 1(a), with the top width of about 103 µm and the bottom width of about 42 µm. The average groove depth is about 1.16 µm.

As shown in Figs. 1(b) and 1(c), the doubly periodic sample has alternating groove separations: $a_1 = 149$ µm and $a_2 = 300$ µm, which are also measured from falling edge to falling edge. There are 40 grooves in total, spanning a length of 8.7 mm. Each groove is about 65 µm wide across the top and 27 µm wide across the bottom. The slope of the falling and rising edges of the grooves is about 0.054, according to Fig. 1(c). The average groove depth is about 0.84 µm. This sample could be equivalently defined by a "super-lattice" of grooves, where two lattices, each with a lattice constant $a_3 = a_1 + a_2$, are offset by a distance equal to $a_1$.

The film thicknesses ($t$) listed in Table 1 were determined through scanning electron microscopy (SEM) measurements using a ThermoFisher "Scios 2 DualBeam" system. The saturation induction ($4\pi M_s$) values were determined through the numerical fitting of the experimentally measured spin-wave dispersion curves to the theoretical dispersion relation. Such fitting is described in Section III.

## III. Experiment and Analysis Approaches

The spin-wave experiments on each sample were carried out through the placement of two microstrip transducers on the sample and the measurement of the transmission coefficient $S_{21}$ across the two transducers by a vector network analyzer (VNA), as illustrated schematically in Fig. 2(a). Each transducer is a 50-μm-wide, end-shorted strip line with a nominal impedance of 50 Ω. The separation of the two transducers, denoted by $l$, is less than the full span of the grooves. The sample is magnetized to saturation by an external magnetic field ($H_0$) that is applied in the plane of the YIG film or the magnonic crystal and is perpendicular to the sample length direction. This field configuration supports the propagation of magnetostatic surface waves (MSSWs) within the sample. For all the data presented below, the field direction was set to ensure that spin waves travel on the top surfaces of the samples. This means that, for the magnonic crystal samples, the waves propagate along the YIG surfaces etched with grooves, rather than the surfaces interfacing with the GGG substrates. Reversing the field direction would result in spin-wave propagation along the surfaces interfacing with the GGG substrates, consistent with the chirality of the MSSWs.[11,12,13] This reversed configuration produces notably weaker and noisier $S_{21}$ responses.

The amplitude and phase of the complex transmission coefficient $S_{21}$ are both measured. The phase ($\phi$) measured as a function of frequency ($f$) can be expressed as $\phi(f) = k(f)l + \phi_0$, where $k$ is the wavenumber of the spin wave, $k(f)l$ denotes the phase of the spin wave accumulated during its propagation from the excitation transducer to the detection transducer, and $\phi_0$ is the phase associated with the electric components, such as the microwave cables and the connectors. As such, the measurement of $\phi(f)$ allows for the determination of the dispersion $k(f)$ of the spin wave as

$$k(f) = \frac{\phi(f) - \phi_0}{l}. \tag{1}$$

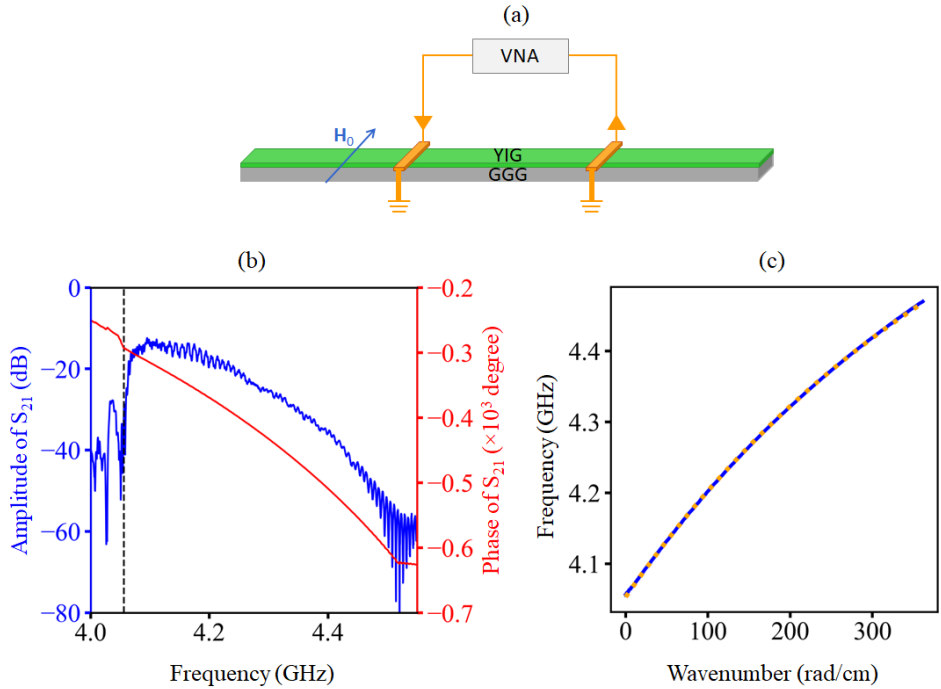

Fig. 2. Spin waves in a uniform YIG film strip. (a) Experimental configuration. (b) Transmission coefficient $S_{21}$ of a transducer/YIG/transducer structure measured under a magnetic field of $H_0 = 782$ Oe applied in the film plane and perpendicular to the YIG strip length. The blue and red profiles show the amplitude and phase of $S_{21}$, respectively, as a function of frequency. The dashed black line marks the FMR frequency. (b) Dispersion relation of the spin wave. The solid blue curve shows the dispersion obtained from the phase profile in (a), while the dotted orange curve is a numerical fit to Eq. (2).

Note that the wavenumber is taken to be positive, which may require taking the absolute value of the numerator of Eq. (1) depending on how the VNA measures the phase.

The theoretical dispersion for an MSSW in a thin film takes the form[11,14]

$$f(k) = \frac{|\gamma|}{2\pi}\sqrt{H_0(H_0 + 4\pi M_s) + \frac{1}{4}(4\pi M_s)^2(1 - e^{-2kt})} \qquad (2)$$

where $\frac{|\gamma|}{2\pi} = 2.8$ MHz/Oe is the absolute gyromagnetic ratio, $4\pi M_s$ is the saturation induction, and $t$ is the film thickness. According to Eq. (2), the $f(k)$ curve increases monotonically between the bounds $f_B = \frac{|\gamma|}{2\pi}\sqrt{H_0(H_0 + 4\pi M_s)}$ and $f_S = \frac{|\gamma|}{2\pi}(H_0 + \frac{1}{2}4\pi M_s)$, where $f_B$ and $f_S$ are the frequencies in the limits of $k = 0$ and $k \to \infty$, respectively. Note that $f_B$ is the Kittel equation describing the ferromagnetic resonance (FMR) frequency of the film under an in-plane field. In practice, with an increase of $f$, the spin-wave signal measured by the detection transducer will have decayed to the noise floor at a frequency below $f_S$. The primary source of this decay is associated with the spin-wave group velocity defined as $v_g = \frac{d(2\pi f)}{dk}$. In a simplified picture, one can express the spin-wave decay factor as $e^{-\eta l/v_g}$, where the decay rate $\eta$ scales with the Gilbert damping constant: a dimensionless parameter measuring magnetic relaxation in a material. As $f$ is increased towards $f_S$ or $k$ approaches infinity, $v_g$ approaches zero according to Eq. (2) and $e^{-\eta l/v_g}$ therefore also approaches zero. A second source is from the filtering effect of the microstrip transducers: the excitation and detection efficiencies of the transducers are relatively high at low $k$ but decrease with an increase in $k$.[15]

In this work, Equation (1) is used to determine the experimental dispersion curve, and Equation (2) is then used to fit it. The numerical fitting yields an effective $4\pi M_s$ value for the sample. Such values are presented in the last row of Table 1.

To demonstrate this analysis process, Figures 2(b) and 2(c) show experimental data and numerical fitting for the uniform YIG strip. Figure 2(b) presents the $S_{21}$ data measured as a function of $f$ under a magnetic field of $H_0 = 782$ Oe. The blue and red profiles show the amplitude and phase of $S_{21}$, respectively. The vertical dashed line marks the FMR frequency, indicated by $f_B$ which is defined above. One can see that as $f$ increases, the $S_{21}$ amplitude decays, as explained earlier.

Figure 2(c) shows the spin-wave dispersion. The solid blue curve is the experimental dispersion curve obtained from the phase profile in Fig. 2(b) by the use of Eq. (1). One can see that $f$ increases with $k$ monotonically with a negative curvature, as expected.[14] The dotted orange curve is a fit to Eq. (2) using Python and the "curve_fit" method in the SciPy library. The fitting yields an effective saturation induction of $4\pi M_s = 1897.7 \pm 0.1$ G and a film thickness of $t = 9.607 \pm 0.003$ µm. The thickness value here is slightly smaller than the value in Table 1 (10.5 µm) which was obtained from the SEM measurement. This difference is likely due to the uncertainty of the thickness measurement associated with the quality of the SEM image. If the film thickness is kept constant at 10.5 µm and $4\pi M_s$ is the only free parameter, the fitting yields a similar curve with $4\pi M_s = 1881 \pm 0.2$ G which is slightly smaller and suggests that the thickness parameter is relatively insensitive to the order of a micrometer.

## IV. Experimental Results

Figures 3 and 4 present the main experimental data of this work. Figure 3 gives the $S_{21}$ data in (a) and (b), the dispersion data in (c), and the Bragg wavenumbers in (d) for the singly periodic magnonic crystal. The format of Fig. 3(a) is the same as that of Fig. 2(b) – the blue and red profiles show the amplitude and phase, respectively, of $S_{21}$ measured as a function of $f$. Figure 3(b) presents the same data as Fig. 3(a) but provides a zoomed-in view over a narrow frequency range around the third dip in Fig. 3(a). The format of Fig. 3(c) is the same as that of Fig. 2(c) – the blue solid curve shows the dispersion obtained from the phase data in Fig. 3(a) using Eq. (1), while the orange dotted curve shows the numerical fit to Eq. (2). The insert in Fig. 3(c) gives a zoomed-in view of the dispersion near the frequency of the third dip in Fig. 3(a). In contrast to the data presented in Figs. 2(b) and 2(c) for the uniform YIG strip, the data in Figs. 3(a), 3(b), and 3(c) show characteristic dips in the amplitude profile, as indicated by the vertical arrows, and phase and dispersion jumps exactly at the center frequencies of the dips.

The transmission dips and the corresponding phase and dispersion jumps shown in Figs. 3(a), 3(b), and 3(c) are indications of the formation of the band gaps for the spin waves in the magnonic crystal. The wavenumbers associated with the centers of the band gaps satisfy the Bragg condition

$$k_n a = n\pi \tag{3}$$

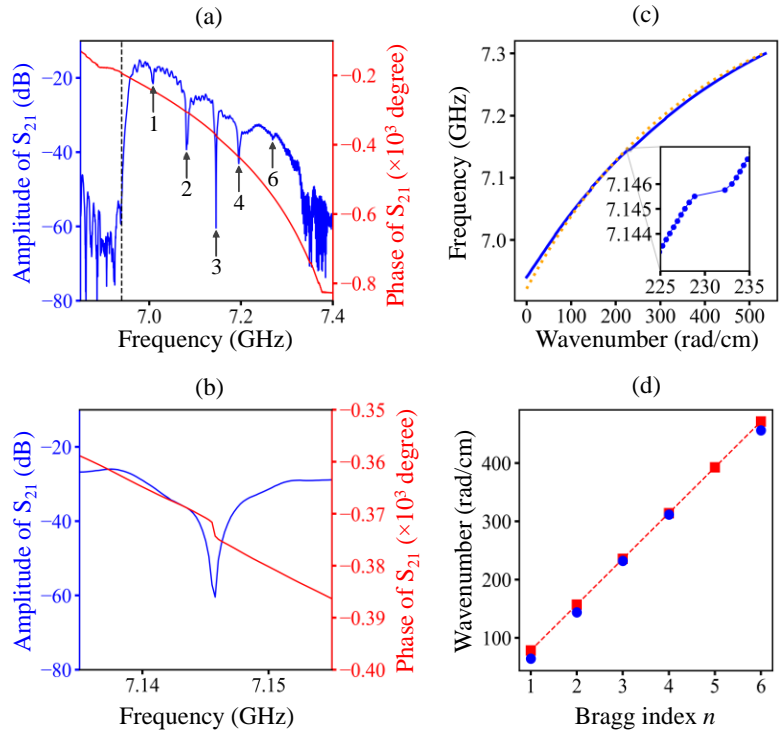

Fig. 3. Spin waves in a singly periodic magnonic crystal, with a period of $a = 400$ μm. (a) Transmission coefficient $S_{21}$ of the magnonic crystal measured under a field of $H_0 = 1690$ Oe applied in the sample plane and perpendicular to the sample length direction. The blue and red profiles show the amplitude and phase of $S_{21}$, respectively, as a function of frequency. The dashed black line marks the FMR frequency. The short vertical arrows point to the centers of the band gaps. (b) Zoomed-in view of (a) near the third band gap. (c) Dispersion curve. The blue curve shows the measured dispersion response while the orange dotted curve is a numerical fit. (d) Measured Bragg wavenumbers (blue points) compared to calculated Bragg wavenumbers (red squares).

where $a = 400$ µm is the groove period or lattice constant of the magnonic crystal and $n = 1, 2, ...$ is the Bragg index. $k_n$ in Eq. (3) is referred to as Bragg wavenumber in the following discussion.

Based on the experimental dispersion curve and the transmission dips, five Bragg wavenumbers are identified, which are shown by the blue points in Fig. 3(d). According to Eq. (3), one can readily calculate the Bragg wavenumbers for $n = 1, 2, ..., 6$. These calculated wavenumbers are presented as the red points in Fig. 3(d). One can see a good agreement between the measured and calculated Bragg wavenumbers. This agreement justifies that the experimentally observed transmission dips are indeed the indications of the band gaps. Further justifications are presented in Section V. However, the comparison in Fig. 3(d) clearly indicates the absence of $k_5$ in the experimental data. This is discussed briefly in Section V. Note that a fitting of the experimental data point in Fig. 3(d) against Eq. (3) yields $a = (378 \pm 17)$ µm, which is slightly smaller than the period (400±4 µm) determined through the surface profile measurements.

Further, the data in Figs. 3(a) and 3(b) also show that the transmission dips appear similar in the sense that they have similar shapes and linewidths. They do exhibit different depths, as explained in the following. In an idealized single-periodic magnonic crystal, one might expect all Bragg modes to exhibit similar or uniform reflection characteristics, giving rise to similar transmission dips. However, in real systems, the reflection behavior of each mode is influenced by its unique spin-wave profile and boundary condition configuration. These factors lead to variations in the depth of transmission dips. As shown in Fig. 3(a), the third dip exhibits the greatest depth. This pronounced feature can be attributed to the $n = 3$ mode experiencing the strongest constructive interference; The alignment of its nodes and antinodes with the grooves effectively enhances reflection at the corresponding frequency of the third dip. Additionally, wavenumber-dependent properties, such as group velocity and two-magnon scattering, further disrupt the theoretical uniformity of reflection features, causing certain modes to stand out more prominently than others.

Turn now to the data presented in Fig. 4 for the doubly periodic magnonic crystal. To ease the comparison, the data in Fig. 4 are presented in the same format as those in Fig. 3. When compared to the $S_{21}$ data in Fig. 3, the data in Fig. 4 show two major differences: (1) the third dip from the left is much broader and (2) the sixth dip is significantly more pronounced. Further, when compared with other dips in the same transmission profile, one can see that (1) the third dip is the strongest and (2) the sixth dip is stronger than both the fourth and fifth dips. These observations together suggest that every third dip represents a special condition when compared with the other dips.

Figure 4(d) presents the experimental Bragg wavenumbers in blue points and three sets of calculated Bragg wavenumbers in red squares. From top to bottom, those three sets of Bragg wavenumbers were calculated for lattice constants $a_1 = 149$ µm, $a_2 = 300$ µm, and $a_3 = a_1 + a_2 = 449$ µm, respectively, as indicated. The comparison of the measured and calculated Bragg wavenumbers gives two important results. First, there is a good agreement between the experimental values and the values calculated with a lattice constant of $a_3 = a_1 + a_2 = 449$ µm. This means that the six dips in the transmission profile correspond to six band gaps that are formed due to the Bragg reflection at $k_n(a_1 + a_2) = n\pi$ where $n = 1, 2, \cdots, 6$. Second, for the third dip, the Bragg wavenumber simultaneously satisfies three Bragg conditions, as indicated by the horizontal dashed line. Those three conditions are $k_3 a_3 = 3\pi$ for the lattice constant $a_3 = 449$ μm, $k_2 a_2 = 2\pi$ for the lattice constant $a_2 = 300$ μm, and $k_1 a_1 = \pi$ for the lattice constant $a_1 = 149$ μm. For the sixth dip, the Bragg wavenumber also simultaneously satisfies three Bragg conditions, as indicated by the horizontal dotted line. This result indicates that the formation of every third band gap involves triple Bragg reflections. This clearly explains why the third and sixth dips are more pronounced.

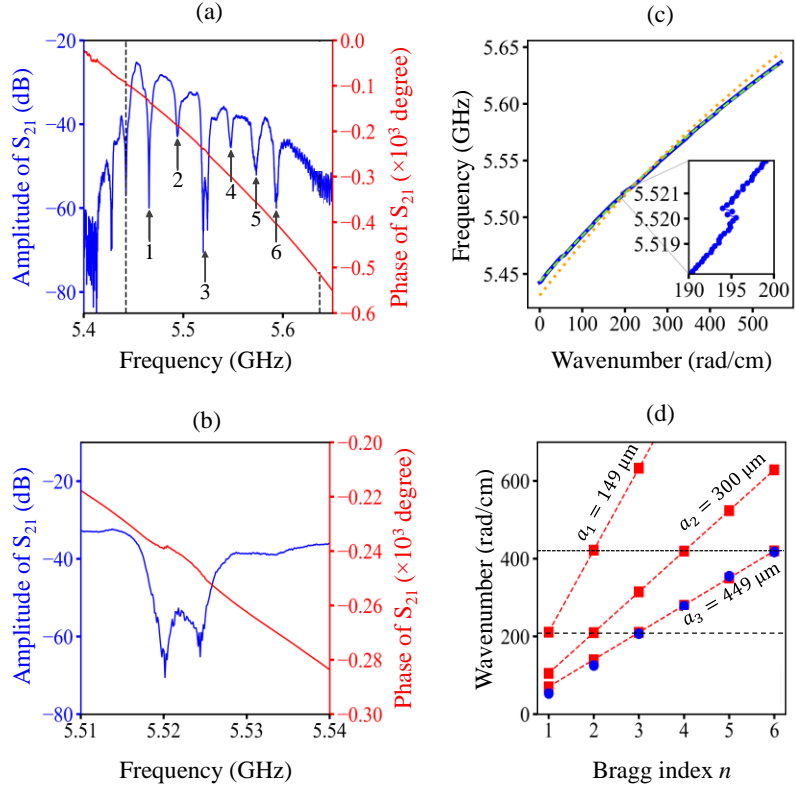

Fig. 4. Spin waves in a doubly periodic magnonic crystal, with $a_1 = 149$ μm and $a_2 = 300$ μm. (a) Transmission coefficient $S_{21}$ of the magnonic crystal measured under a field of $H_0 = 1240$ Oe applied in the sample plane and perpendicular to the sample length direction. The blue and red profiles show the amplitude and phase of $S_{21}$, respectively, as a function of frequency. The dashed black line marks the FMR frequency. The vertical arrows point to the centers of the band gaps. (b) Zoomed-in view of (a) near the third band gap. (c) Dispersion curve. The solid curve shows the measured dispersion response while the dotted and dashed curves are numerical fits. (d) Measured Bragg wavenumbers (blue points) compared to calculated Bragg wavenumbers (red squares).

Three notes should be made about the experimental data presented above. First, the spin waves shown by the data in Figs. 2, 3, and 4 have different frequencies which are due to the use of different magnetic field strengths ($H_0$) during the measurements. Adjusting the field strength would shift the $S_{21}$ responses along the frequency axis, but this would not affect the qualitative results described above.

Second, for $f < f_B$, the $S_{21}$ amplitude vs. frequency profiles reveal some non-trivial responses above the noise level. These responses are particularly noticeable in Fig. 2(b) and Fig. 4(a). The presence of such responses is associated with the fact that the YIG waveguide in this study is not exactly one-dimensional and spin waves along the field direction are therefore not completely absent. Spin waves propagating along an in-plane field are classified as backward volume waves whose frequencies are below $f_B$.[11,14] Further, in the $f < f_B$ regime shown in Fig. 4(a), the $S_{21}$ amplitude also shows a dip at about 5.43 GHz. This dip may be associated with one of the width modes across the YIG sample width, and one may confirm this association by examining how the dip varies as the field is rotated in-plane.

Third, the orange dotted curves in Fig. 3(c) and Fig. 4(c) show a noticeable mismatch with the experimental dispersion curves. This discrepancy arises because the fitting process treated the film thickness $t$ as a constant and varied only the saturation induction $4\pi M_s$ as the fitting parameter. The $t$ values used were derived from the SEM measurements, as mentioned earlier, and the fitting-yielded $4\pi M_s$ values are given in Table 1. However, allowing both $t$ and $4\pi M_s$ to vary as free parameters results in nearly perfect fits to the experimental dispersion curves. This improved fit is demonstrated by the green dashed curve presented in Fig. 4(c). The fitting yielded $t$ = 3.6 µm, which is smaller than the value (4.2 µm) from the SEM measurement, and $4\pi M_s$= 1801 G, which is slightly larger than the value (1790 G) obtained with $4\pi M_s$ as the sole fitting parameter. The notable difference in film thickness likely stems from errors in the SEM-based measurement, though this does not significantly impact the analysis in this work since the focus is on surface spin waves rather than volume spin waves.

## V. Numerical Simulations

To further support the above explanation of the experimental observations, numerical simulations were carried out in Python using the transfer matrix method (TMM), a well-established technique for analyzing spin-wave propagation.[6,10,16,17,18] In this approach, a transfer matrix $A$ transforms the state of the spin wave across a potential interface by enforcing the continuity of the wavefunction and its derivative:

$$\begin{pmatrix}\psi_2\\\psi_{2'}\end{pmatrix} = A_2^1 \begin{pmatrix}\psi_1\\\psi_{1'}\end{pmatrix}, \tag{4}$$

where $\psi_1$ and $\psi_2$ represent the wavefunction within the regions on either side of the interface and the primed variable $\psi'$ denotes the spatial derivative. The transfer matrix $A_2^1$ is explicitly constructed from the wavenumbers $k_1$ and $k_2$ within the respective regions by taking the assertion of continuity in Eq. (4) and rearranging for the outgoing signal in terms of the incoming signal.[19] The result is

$$A_2^1 = \begin{bmatrix}\frac{1}{2}\left(1+\frac{k_1}{k_2}\right) & \frac{1}{2}\left(1-\frac{k_1}{k_2}\right)\\\frac{1}{2}\left(1-\frac{k_1}{k_2}\right) & \frac{1}{2}\left(1+\frac{k_1}{k_2}\right)\end{bmatrix}. \tag{5}$$

To simulate the experimental configuration composed of multiple interfaces, the full potential profile of the magnonic crystal samples is discretized into regions each associated with its own transfer matrix. The wavenumber within each region is a function of the sample thickness at that position $x$:

$$k(x) = k_0 \frac{t}{t-D(x)}, \tag{6}$$

where $t$ is the uniform film thickness and $D(x)$ is the groove depth. The flat regions corresponding to the uniform sections and the bottoms of grooves are represented by individual transfer matrices, while the sloping regions in between are divided into $N$ potential steps. Convergence is ensured by choosing $N$ sufficiently large such that the difference between results for $N$ and $N/2$ is less than $10^{-10}$.

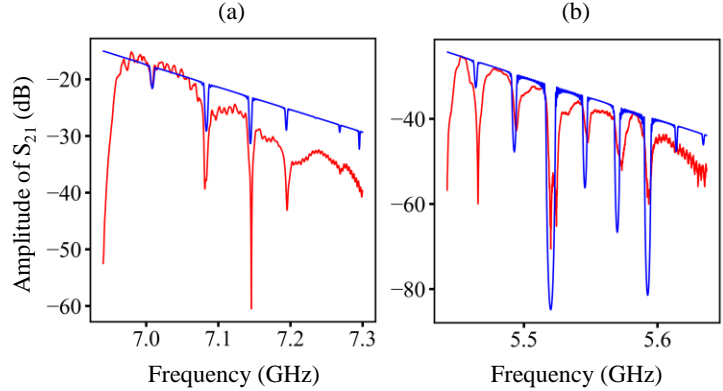

Fig. 5. Comparison of experimental (red) and simulated (blue) transmission profiles in (a) the singly periodic magnonic crystal and (b) the doubly periodic magnonic crystal.

The simulations use the measured sample properties presented in Table 1 and the applied field strengths ($H_0$) as used in Section IV. For each initial wave number, the transfer matrix for each region is computed and the transmittance and reflectance are derived. This process is repeated for all wavenumbers corresponding to the frequencies measured in the experiment. The transfer matrix Eq. (5) does not account for any lossy interactions, so the simulation takes spin-wave decay into account with a factor $e^{-ak+b}$ to represent the physical decay term $e^{-\eta l/v_g}$ discussed in Section III and scale the simulation amplitude to the experiment. This factor provides a better visual comparison to the experimental results without affecting the location of the Bragg reflections, but it was not rigorously determined.

Figure 5 compares the simulated transmission profiles (blue) with the experimental profiles (red). The profiles in (a) and (b) are for the singly and doubly periodic magnonic crystals, respectively. The comparison gives three important results. First, the transmission dips in the simulated profiles align with those in the experimental profiles. Second, for the profiles in Fig. 5(a), the relative strengths (depths) of the simulated dips agree with those of the experimental dips. These two results together clearly show the accuracy of the numerical simulations. Finally, the numerical profiles in Fig. 5(b) evidently confirm the experimental observation that for the doubly periodic magnonic crystal, the third and sixth dips are more pronounced than other dips.

To confirm the interpretation of the strong third and sixth dips given in the last section, Figure 6 compares two numerical transmission profiles. The blue profile, which is the same as in Fig. 5(b), is the simulation of the doubly periodic magnonic crystal using the dimensions as measured and shown previously, while the red profile is the simulation of a magnonic crystal with a single period equal to $a_1 + a_2 = 449$ μm. The simulations used the same magnetic field as in the experiment, which is $H_0 = 1240$ Oe. These simulations explicitly show the effect on the transmission dips due to the second period and thereby confirm that the pronounced dips result from multiple Bragg reflections associated with the double periodicity.

Additionally, the numerical profile in Fig. 5(a) reveals a very tiny dip at about 7.235 GHz, which is the fifth dip from the left. This explains why the fifth dip is not visible in the transmission profile in Fig. 3(a) and the $k_5$ point is missing in Fig. 3(d). As discussed earlier, the reflection behavior of each mode is influenced by its unique spin-wave profile and boundary condition configuration. For the $n = 5$ mode, the specific configuration results in a minimal transmission dip. This finding has been verified by additional simulations. For instance, increasing the groove width by about 50% significantly enhances the depth of the fifth dip, making it even deeper than the fourth dip, while maintaining the same dip locations.

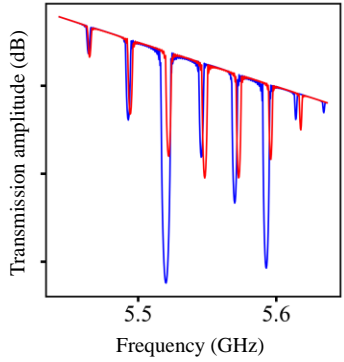

Fig. 6. Simulated transmission profiles of the doubly periodic magnonic crystal with $a_1 = 149$ μm and $a_2 = 300$ μm (blue) and a magnonic crystal with a single periodicity $a_3 = a_1 + a_2$ (red).

## VI. Final Remarks

In conclusion, this work explored the effects of double periodicity on the band gaps of spin waves in a 1D magnonic crystal. Transverse line grooves were chemically etched into a long and narrow YIG thin film strip to create a magnonic crystal with alternating lattice constants $a_1 = 149$ μm and $a_2 = 300$ μm. The propagation of spin waves in this doubly periodic magnonic crystal was measured experimentally using a VNA and simulated numerically using a transfer matrix method. The experimental and numerical results show that the third and sixth band gaps are more pronounced than the other band gaps; the underlying physics is that the wavenumbers associated with those band gaps satisfy simultaneously triple Bragg conditions for periods $a_1 = 149$ μm, $a_2 = 300$ μm, and $a_3 = a_1 + a_2 = 449$ μm. These results indicate that the introduction of a secondary periodicity into a magnonic crystal represents an effective strategy for tailoring band gap properties in magnonic crystals. Future work that explores in-gap magnonic states or nonlinear spin-wave dynamics in similar doubly periodic magnonic crystals is of great interest. It is also of great interest to examine spin-wave dynamics in doubly periodic magnonic crystals where the two periods $a_1$ and $a_2$ are incommensurable.

It is worth noting that finite-size effects and mode hybridizations might also play a role in enhancing the third and sixth band gaps. However, these contributions are believed to be minimal for the following reasons. First, multiple reflections of spin waves from the two ends of the YIG waveguides could produce a Fabry–Pérot-like effect, which might enhance the observed band gaps. Nonetheless, this effect is expected to be weak, as the YIG sample lengths (e.g., 22.6 mm for the doubly periodic sample) in this study are substantially larger than the transducer separations (e.g., 7.4 mm for the doubly periodic sample). As a result, the spin waves reflected from the sample ends undergo significant attenuation, especially when considering the non-reciprocal nature of the MSSWs. Additionally, the numerical simulations in this study were deliberately designed to exclude explicit boundary reflections, effectively mimicking an infinite magnonic crystal without a Fabry–Pérot cavity. These boundary-free simulations reproduced the enhanced third and sixth dips at precisely the same frequencies observed experimentally. This strong correspondence suggests that the observed band-gap enhancements are an intrinsic consequence of the double periodicity rather than artifacts of finite sample length.

The geometry and magnetic field configuration used in this study predominantly excited the fundamental surface wave mode, while higher-order modes (i.e., thickness modes) were minimally populated during measurements. As a result, there was no significant mode hybridization with higher-order modes. Additionally, each etched groove in the magnonic crystal could potentially host localized states or defect modes. If such modes are coupled strongly with traveling spin waves, they might produce additional gaps or sidebands. However, the primary transmission dips observed coincide with the triple Bragg condition predicted by the TMM model, indicating that these features are not due to localized modes.

Finally, a note on the phase analysis is warranted. In this work, the extraction of spin-wave dispersion from the experimentally measured phase data assumed that the spin waves in the magnonic crystals behave as plane waves, even though the periodic structure inherently supports Bloch-type spin waves. Two factors justify this approximation. First, the measured $S_{21}$ phase primarily reflects the phase accumulation between the excitation and detection transducers. While the grooves introduce periodic perturbations, the net phase shift measured by the VNA effectively represents an average of the local wavevector along the entire propagation path. Consequently, the total phase accumulation can be approximated by an effective wavevector. Second, while the TMM model accurately describes Bloch-like wave propagation through a discretized stack of grooves, the calculated band-gap frequencies align closely with those inferred from the simpler plane-wave phase data. This alignment suggests that small deviations from the plane-wave approximation do not significantly affect the principal spin-wave dispersion in the magnonic crystals studied here.

Lastly, it should be noted that the YIG films in the three samples used in this study have varying thicknesses. However, these differences do not substantially impact the main conclusions, as all measurements were performed in a surface wave configuration. Thus, the dynamics examined are primarily confined to the top surfaces of the films, rather than spanning their entire thickness.

**Acknowledgements:** This work was supported in part by the US National Science Foundation under Grant DMR-2420266.